\documentclass[journal=nalefd,manuscript=letter]{achemso}

\usepackage{amsmath}
\usepackage{hyperref}
\usepackage{xfrac}
\usepackage{array}
\usepackage{notes2bib}
\graphicspath{ {Figures/} }
\title{Purification of single photons by temporal heralding of quantum dot sources}

\author{Hamza Abudayyeh}%
\author{Boaz Lubotzky}
 \affiliation{%
Racah Institute for Physics, the Applied Physics Department and the Center for Nanoscience and Nanotechnology, The Hebrew University of Jerusalem, Jerusalem 9190401, Israel}
\author{Somak Majumder}
\author{Jennifer A. Hollingsworth}
\affiliation{Materials Physics \& Applications Division: Center for Integrated Nanotechnologies, Los Alamos National Laboratory, Los Alamos, New Mexico 87545, United States}

\author{Ronen Rapaport}
 \email{ronenr@phys.huji.ac.il}
 \affiliation{%
Racah Institute for Physics and Center for Nanoscience and Nanotechnology, The Hebrew University of Jerusalem, Jerusalem 9190401, Israel}

\begin{document}

\maketitle
% ABSTRACT
\begin{abstract}
Efficient, high rate photon sources with high single photon purity are essential ingredients for quantum technologies. Single photon sources based on solid state emitters such as quantum dots are very advantageous for integrated photonic circuits, but they can suffer from a high two-photon emission probability, which in cases of non-cryogenic environment cannot be spectrally filtered.  
Here we propose two \textit{temporal} purification-by-heralding methods for using a two photon emission process to yield highly pure and efficient single photon emission, bypassing the inherent problem of spectrally overlapping bi-photon emission at elevated temperatures.
We experimentally emulate their feasibility on the emission from a single nanocrystal quantum dot at room temperature, exhibiting single photon purities exceeding 99.5\%, without a significant loss of single photon efficiency. 
By utilizing these methods with commercially available components, nanocrystal quantum dots  can be made realistic high-quality room-temperature sources for single photons. These approaches can also be applied for any indeterministic source of spectrally broadband photon pairs.

\end{abstract}

%%%%%%%%%%%%%%%%%%%%%%%%%%%%%%%%%%%%%%%%%%%%%%%%%%%%%%%%%%%%%%%%%%%%%%
%%                           MAIN TEXT    							%%
%%%%%%%%%%%%%%%%%%%%%%%%%%%%%%%%%%%%%%%%%%%%%%%%%%%%%%%%%%%%%%%%%%%%%%
%% INTRODUCTION
\section{Introduction:}
A single photon source (SPS) is proposed as a crucial resource for several major applications in quantum technologies \cite{Lounis2005,Eisaman2011InvitedDetectors,Aharonovich2016Solid-stateEmitters}.
The required performance of such a source depends on the application it would be used for \cite{Aharonovich2016Solid-stateEmitters}, however generally most applications require stable, compact, high-rate and high-single-photon-purity SPSs. 
%A few types of other light sources, which do not strictly emit single photons were developed and are being used frequently. These include faint lasers or 
The most widely used resource for single photons comes from low efficiency down-converted photon-pairs from non-linear processes in solids \cite{PaulG.Kwiat1995NewPairs,Yan2015GenerationWaveguide,Ngah2015Ultra-fastTechnology,Joshi2018FrequencySources}. 
Recently however, research has been focused on developing more efficient solid state SPSs such as semiconductor self assembled quantum dots (QDs) \cite{Ding2016,Somaschi2016NearState,Schlehahn2016,Daveau2017EfficientWaveguide,Liu2017HighPhotons}, chemically synthesized semiconductor nanocrystal quantum dots (NQDs) \cite{Lounis2000PhotonFluorescence,Michler2000QuantumTemperature,Chen2008GiantBlinking,Bidault2016PicosecondTemperature,Hoang2016UltrafastNanocavities,Matsuzaki2017StrongAntenna},  color centers in crystals \cite{Aharonovich2014DiamondNanophotonics,Schroder2016QuantumInvited,Castelletto2014ASource}, carbon nanotubes \cite{He2017TunableNanotubes,Ishii2017Room-TemperatureNanotubes}, and defect states in two-dimensional materials \cite{Chakraborty2015Voltage-controlledSemiconductor,Grosso2017TunableNitride} among others. 

One crucial requirement from an SPS is a very high single photon purity, i.e., a high probability that an emission event contains a single photon. 
Unfortunately most solid-state SPSs, especially room-temperature sources, do not inherently have sufficiently high purities. 
QD-based SPSs which operate at cryogenic temperatures, for example, have resolvable multi-exciton spectral lines.
Therefore multi-exciton emission can be spectrally filtered leaving only single photon emission from the single exciton state (X). 
However, the necessity of a  cryogenic work environment is limiting technologies, and semiconductor QD sources working at elevated temperatures have only been achieved with either wide bandgap QDs \cite{Arita2017UltracleanDot}, or using the smaller NQDs \cite{Lounis2000PhotonFluorescence,Michler2000QuantumTemperature}.  
Yet, operating at elevated temperatures introduces a significant limitation in maintaining both high purity together with a high source efficiency due to the spectral overlap between the single exciton (X) and bi-exciton (BX) emission.
The severity of BX emission increases significantly when the Auger recombination processes \cite{Efros2016OriginDots} are suppressed to increase QD stability or when NQDs are radiatively enhanced to achieve higher rates \cite{Park2013Single-NanocrystalTemperature,Wang2015CorrelatedPathways,Dey2016PlasmonicDots}. 
This detrimental problem of reduced purity is rather inherent for such elevated temperature SPSs, and no solution to increase purity without a significant sacrifice in terms of brightness has been successfully proposed or implemented to the best of our knowledge. 

Here we propose and demonstrate two \textit{temporal} purification-by-heralding methods for using a two photon emission process to yield highly pure and efficient single photon emission, thus bypassing the inherent problem of spectrally overlapping bi-photon emission.
We then experimentally emulate these temporal purification methods on the emission from a single NQD that is particularly prone to efficient BX emission \cite{Mangum2014InfluenceNanocrystals}, and show that employing such techniques enable reaching single photon purities exceeding 99.5\%, without a significant loss of single photon efficiency.
%These methods can be applied for any indeterministic source of photon pairs.

%% THEORETICAL FRAMEWORK 
\section{Purification by Temporal Heralding:}
\begin{figure}[ht!]
\centering{\includegraphics{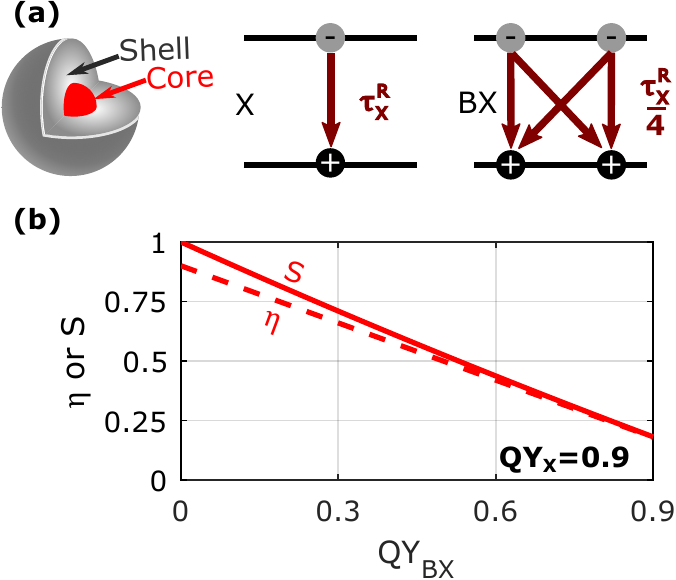}}
\caption{(a) Illustration of a spherical nanocrystal quantum dot displaying the radiative recombination pathways available for the X and BX state. (b) Single photon purity (solid line) and efficiency (dashed line) for a quantum dot as a function of the BX quantum yield.  }
\label{fig:QDandTGF}
\end{figure}

Chemically synthesized NQDs (see Fig. \ref{fig:QDandTGF}a) are an excellent exemplary system to demonstrate and apply the temporal photon purification and heralding concepts.
NQDs are very promising for integrated and efficient SPSs at room temperature \cite{Brokmann2004ColloidalSources,Livneh2015EfficientNanoantenna,Livneh2016,Lin2017Electrically-drivenTemperature,Chandrasekaran2017NearlyDots}.
Unfortunately, it turns out that the photostability (in terms of photo-blinking) of an NQD can be intimately related to its BX quantum yield (QY$_{BX}$) defined as the probability that the BX would radiatively decay to the X state.
A key mechanism responsible for photo-blinking involves Auger-mediated non-radiative recombination of charged excitons \cite{Efros2016OriginDots}.
Thus, as the stability of the source against photo-blinking increases,  QY$_{BX}$ also usually increases due to the suppression of Auger recombination processes \cite{Efros2016OriginDots}, leading to a dramatic drop of both the source efficiency ($\eta=P_1$) and purity ($S=\sfrac{P_1}{(P_1+P_2)}$), where $P_1$ and $P_2$ are the probability of emitting one and two photons respectively (see Fig. \ref{fig:QDandTGF}b). 
Higher order multi-exciton emission is usually negligible in these quantum dots due their strong excitation power dependence and due to the cubic (or nearly cubic) scaling of the Auger recombination rate with exciton multiplicity whereas the radiative decay rate only has a quadratic scaling \cite{Htoon2010HighlyNanocrystals}. 
Furthermore the emission of these multi-excitons can be spectrally separated from the exciton line at room temperature \cite{Htoon2010HighlyNanocrystals}.
Thus NQD typically emits at most two photons per excitation cycle from the BX and X state with probabilities given by the BX and X quantum yields (QY$_{BX}$ and QY$_X$ respectively). 
Notably a similar increase in QY$_{BX}$ has been reported when NQDs were coupled to plasmonic emission rate enhancers \cite{Park2013Single-NanocrystalTemperature,Wang2015CorrelatedPathways,Dey2016PlasmonicDots}, showing that trying to increase the inherent emission rate of an NQD via the Purcell effect  has a detrimental effect on the purity of the source. 
Therefore it seems necessary to develop some photon purification methods that will enable utilizing NQDs as efficient high-quality room temperature SPSs.

Since neither spectral nor polarization based filtering of the BX emission is possible, a different approach is essential. 
One such approach, referred to here as time-gated filtering (TGF), is available in the time domain by using the difference in the lifetimes between the X and BX  to temporally filter out (and discard!) the BX emission in a time-resolved measurement (see Fig. \ref{fig:QDandTGF}a) \cite{Mangum2013DisentanglingExperiments.,Feng2017PurificationDots}.
In this approach however, due to the overlap between the BX and X decays, a very unfavorable  trade-off between the achievable purity and efficiency appears for high BX and X QYs \bibnote[supp]{see supplementary information for further details.}.
For example, in the theoretical limit when $QY_X=QY_{BX}=1$ obtaining a purity of 99.5\% would entail operating at an efficiency of about 25\% even if the collection efficiency of the system, $\alpha$, is unity  \cite{supp}. For currently available high-quality high-rate NQD emitters having $QY_X=0.61,QY_{BX}=0.7$ \cite{Matsuzaki2017StrongAntenna}, the efficiency would only be 8\% for $\alpha=1$.    
Therefore such an approach is not so useful for a practical high-quality SPS which requires both $\eta$ and $S$ to approach unity. 

An alternative approach would be to use the two photon events in a heralded scheme to produce highly pure single photon states.
The simplest possible heralding approach, referred to here as the beam splitter heralded scheme, would be to use a 50:50 beam splitter that would randomly direct the emission either to an idler port or a signal port. 
In a post processing step only signal photons heralded by an idler photon would be chosen, assuring $S=1$ in the signal channel.
This technique however has a maximum theoretical efficiency of $\eta=0.5$ in the ideal case where $QY_{X}=QY_{BX}=1$ and $\alpha=1$, due to the random routing of the beam splitter. 

\begin{figure*}[t]
\centering{\includegraphics[width=\textwidth]{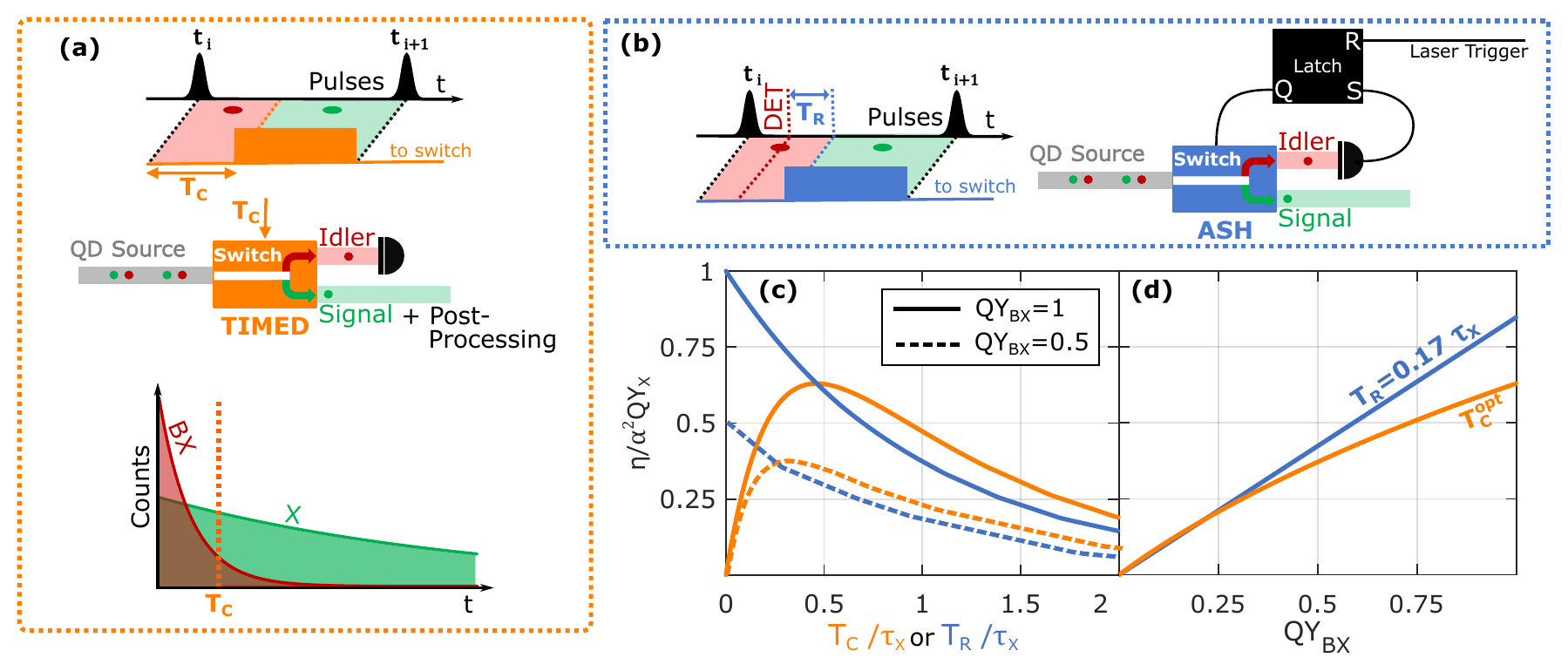}}
\caption{Schematic representations of the TIMe-resolved heraldED (TIMED), and (b) Active Switching Heralded (ASH) techniques proposed in this work. (c) Theoretical calculations of the heralding efficiency as a function of the cutoff time $T_C$ (for TIMED) or the response time $T_R$ (for ASH) normalized by the collection efficiency and the exciton quantum yield. Solid lines correspond to unity BX quantum yield whereas dashed lines correspond to $QY_{BX}=0.5$.  (d) Theoretical normalized heralding efficiency  using optimal parameters as a function of the BX quantum yield assuming unity X quantum yield. }
\label{fig:theoretical}
\end{figure*}

In order to overcome all these limitations we propose two main purification heralded schemes that are illustrated in Fig. \ref{fig:theoretical}a and b. 
\subsection{Passive Heralding:}
The first scheme, called the TIMe-resolved heraldED (TIMED) technique, utilizes the large difference between the BX and X spontaneous radiative lifetimes, and their cascaded emission nature to implement a time gate on an optical switch (see Fig. \ref{fig:theoretical}a).
In this scheme, an optical switch with a fixed time gate will replace the 50:50 beam splitter. 
The switch will be open to the idler port from the beginning of the excitation cycle  up to a fixed cutoff time ($T_C$). 
For $t>T_C$ the switch will direct the photons to the signal port until the the next excitation cycle.\bibnote{In this analysis we assume that the NQD is pumped well above saturation with a short pulse so as to prepare the NQD in the BX state at the beginning of each excitation cycle.}
Here again, only signal photons that were preceded by an idler photon detection would be chosen in a post-processing step.  The single photon efficiency in this scheme is given by \cite{supp}:
\begin{equation}
\begin{split}
\eta_{TIMED}=\alpha^2 & QY_X QY_{BX}\frac{\tau_X}{\tau_X-\tau_{BX}}\times \\ \times&\left[ \exp\left(\sfrac{-T_C}{\tau_X}\right)-\exp\left(\sfrac{-T_C}{\tau_{BX}}\right)\right],
\label{eq: etaTIMED}
\end{split}
\end{equation}
where $\tau_X$ and $\tau_{BX}$ are the X and BX lifetimes respectively.
As seen in Fig. \ref{fig:theoretical}c there is an optimal gate time $T_C^{opt}$  given by:
\begin{equation*}
\frac{T_C^{opt}}{\tau_X}= \left(\frac{\tau_X}{\tau_{BX}}-1\right)^{-1} \ln\left(\frac{\tau_X}{\tau_{BX}}\right). 
\label{eq: Tcopt}
\end{equation*}
The X and BX quantum yields are related to their radiative lifetimes by:
\begin{equation}
\frac{QY_{BX}}{QY_X}=\beta\frac{\tau_{BX}}{\tau_{X}},
\label{eq: QYratio}
\end{equation}
where $\beta$ is a scaling factor which in this case is determined merely by the simple statistical scaling of the optical recombination pathways as a function of the number of excitons \cite{Park2014AugerDots,Matsuzaki2017StrongAntenna}, i.e. $\beta$ = 4. This is shown schematically in Fig. \ref{fig:QDandTGF}a for the X and BX states.
As compared to a standalone NQD (Fig. \ref{fig:QDandTGF}b) the efficiency of TIMED increases with QY$_{BX}$ (Fig. \ref{fig:theoretical}d) illustrating the viability of this technique for high two-photon emission probabilities. 
Furthermore TIMED is a passive scheme requiring no active feedback from the idler port to the switch, which is advantageous due to the simple architecture of the optical circuit. 
Using the above value of $\beta$ and the optimized $T_C^{opt}$ yields a pure source with $S=1$ with a maximum theoretical efficiency $\eta\simeq 0.63$ (again taking $QY_{X}=QY_{BX}=1$, see Fig. \ref{fig:theoretical}c), already better than the simple beam-splitter technique (which can only reach $\eta=0.5$ even in an ideal system).
For currently realized QY values ($QY_X=0.61,QY_{BX}=0.7$) \cite{Matsuzaki2017StrongAntenna} we expect a theoretical efficiency of 26\%, compared to 20\% for the beam-splitter technique. 
This is a significant improvement, yet even better results can be achieved , with an active heralding technique, as is shown next.

\subsection{Active Heralding:}
This maximum efficiency of TIMED is a limitation of passive heralding techniques. To overcome this limitation in $\eta$ without sacrificing purity a more sophisticated solution, the Active Switching Heralded (ASH) technique is suggested (see Fig.\ref{fig:theoretical}b).
In this active switching scheme the switch is open to the idler port from the beginning of the excitation cycle until a photon detection has occurred. 
This detection signal would be forwarded to the switch causing it to route the photons to the signal port until the end of the cycle. 
Realistically, there is a finite time between the arrival of the idler photon to the detector and the switching event which we call the response time ($T_R$) of the system. 
This response time takes into account the rise times of the detector and switch and any optical or electronic delays in the system. 
The heralding efficiency as a function of the response time can be expressed as\cite{supp}:
\begin{equation}
\eta_{ASH}= \alpha^2 QY_X QY_{BX} \exp\left(\sfrac{-T_R}{\tau_X}\right)
\label{eq: etaASH}.
\end{equation}
Notably, it can be seen (Fig.\ref{fig:theoretical}c) that in the ASH scheme, the maximal $\eta$ for the theoretically ideal case $QY_{X}=QY_{BX}=1$ exponentially approaches 100\% (or 43\% for the case $QY_X=0.61,QY_{BX}=0.7$ \cite{Matsuzaki2017StrongAntenna}), as $T_R$ becomes smaller than $\tau_X$. Therefore implementing ASH with very short response gates should approach the ideal SPS scenario.

%% EXPERIMENYAL RESULTS
\section{Experimental Emulation:}
\begin{figure}[t]
\centering{\includegraphics[width=\textwidth]{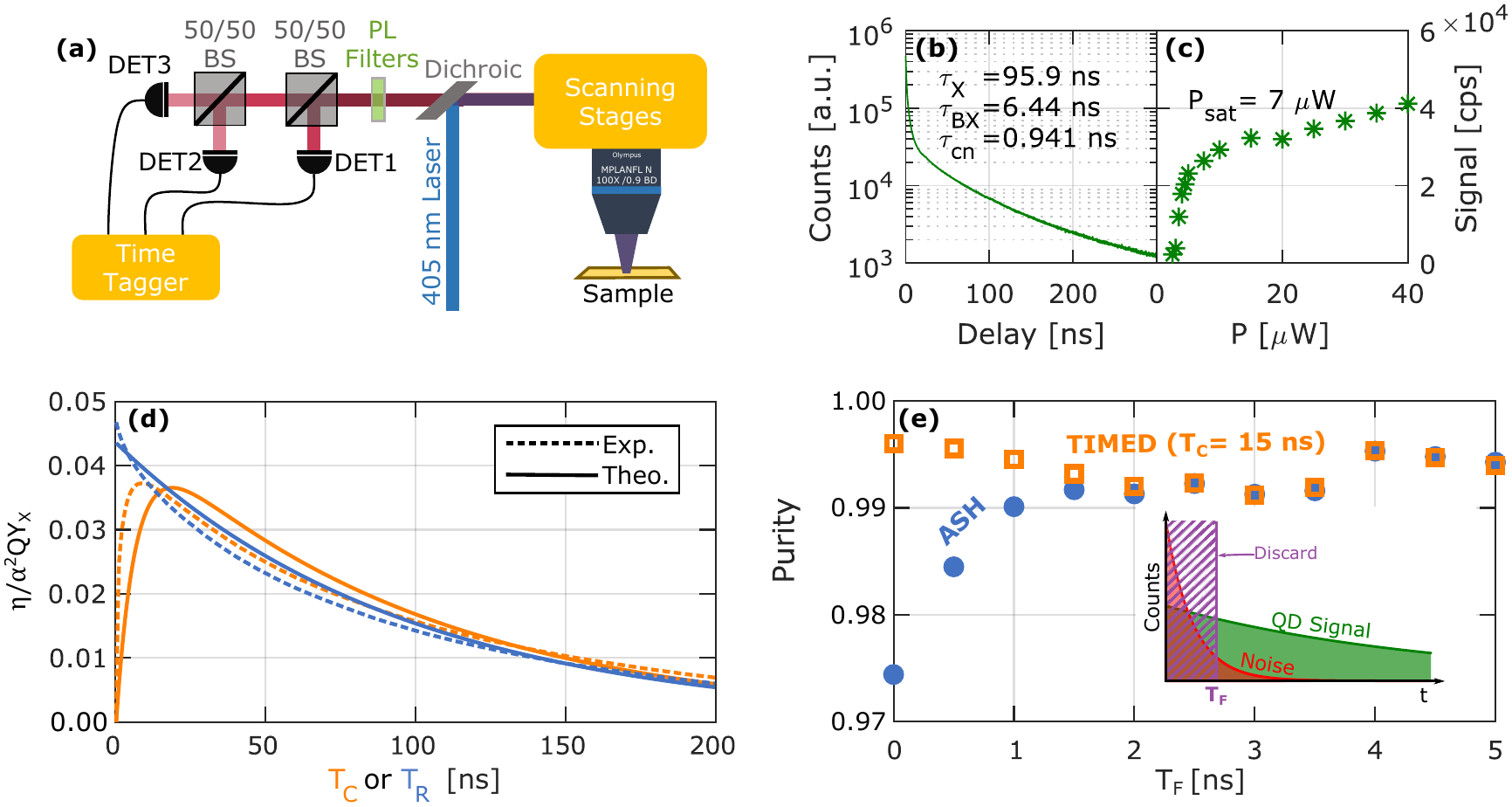}}
\caption{(a) Illustration of the experimental setup used to emulate the TIMED and ASH schemes. (b) Lifetime measurement and (c) power saturation curve for the QD studied in this work.(d) Experimental results (dotted lines) for the efficiency (in units of $\alpha^2QY_X$ in the TIMED and ASH schemes compared to the predicted values (solid lines). (e) The single photon purity as a function of the filtering time which is used to filter out short-time correlated noise illustrated in the inset. }
\label{fig:exp1}
\end{figure}
Rather than physically implementing the setups for the TIMED and ASH protocols, which require building a dedicated setup for each, we emulate the switching on the time-tagged data obtained from unified experimental setup shown in Fig. \ref{fig:exp1}a  
(for details regarding the experimental technique and analysis methods refer to the supplementary material). 
% A core/thick-shell CdSe/CdS NQD \cite{Park2011Near-UnitySpectroscopy} was placed under a 0.9 NA objective and pumped using a 2MHz 405 nm pulsed laser. The NQD emission is directed through two 50/50 beam splitters with three single photon avalanche detectors (Excelitas SPCM-AQRH-14-FC) at their outputs. 
% The counts from the detectors are recorded using a time correlated single photon counting module (Swabian TimeTagger 2.0) operated in time-tagged mode.
% The three detectors are used to measure the purity of the heralded schemes, i.e. once one detector triggers an event we conduct a coincidence measurement on the other two detectors \cite{Note1}.
In such a way we can compare the performance of multiple schemes for the same NQD. 
Fig. \ref{fig:exp1}b displays the lifetime measurement for the NQD under investigation, which was fitted to a three exponential fit yielding the X and BX lifetimes, and a significant short lifetime component which we attribute to a correlated laser-induced noise (hereinafter referred to as correlated noise). 
Fig. \ref{fig:exp1}c displays the saturation curve for the same NQD indicating that at the maximum power used in this experiment we are pumping well above saturation. This is done to ensure a maximal occupation of the BX states.
Using the measured lifetimes and Eq. \ref{eq: QYratio} we find that the ratio between the quantum yields is: $QY_{BX}/QY_X$=0.269. 
With this ratio, the one photon probability, and our calibrated system collection efficiency ($\alpha = 0.088$) we find the value of each of the quantum yields separately to be: $QY_X=0.1729$ and $QY_{BX}=0.0465$ \cite{supp}. 
Prior to any purification protocol we measured the photon purity to be only $S\approx0.9$ for this NQD \cite{supp}.
 
Fig. \ref{fig:exp1}d displays the experimental efficiency for both TIMED and ASH compared to the theoretically expected values (Eqs. \ref{eq: etaTIMED} and \ref{eq: etaASH}) as a function of $T_C$ or $T_R$ respectively. 
% The first 300 ps of data after the laser pulse was ignored to minimize the effects of laser-induced correlated noise and this was also taken into account in calculating the theoretical curves. 
We find a good agreement between the experimental results and the expected values for both schemes,
and the small shift in the position of maximum of the experimental $\eta_{TIMED}$ to a lower $T_C$ is attributed to the effect of the correlated noise. 
We find experimentally that the maximal efficiency for the ASH scheme is $\eta_{ASH}=0.0468 \ \alpha^2 QY_X$, which is in agreement with the experimentally extracted value $QY_{BX} = 0.0465$. Comparing this to Eq. \ref{eq: etaASH} with $T_R=0$ indicates that indeed this efficiency is only limited by the quantum yields of the X and BX, as is theoretically predicted. 

Going to the source purity, theoretically, if the NQD cannot emit more than two-photons per excitation pulse, we expect a unity purity, i.e. $S=1$. Experimentally we expect that in this case the sole source reducing $S$ in the signal port is from noise. 
To check this point we show the purity, $S$, of both schemes as a function of the filtering time $T_F$ after the excitation pulse, 
where all data with $t<T_F$ is ignored. This is done to minimize the contribution of short lifetime laser-induced correlated noise.
For $T_F>1$ns, $S$ reaches a maximum of around 0.995, after filtering out nearly all correlated noise.
We attribute this limit to uncorrelated noise rather to inherent multi-photon ($>2$) emission events from the NQD \cite{supp}.

%% DISCUSSION
\section{Discussion and Outlook:}
\begin{figure*}
\centering{\includegraphics[width=0.5\textwidth]{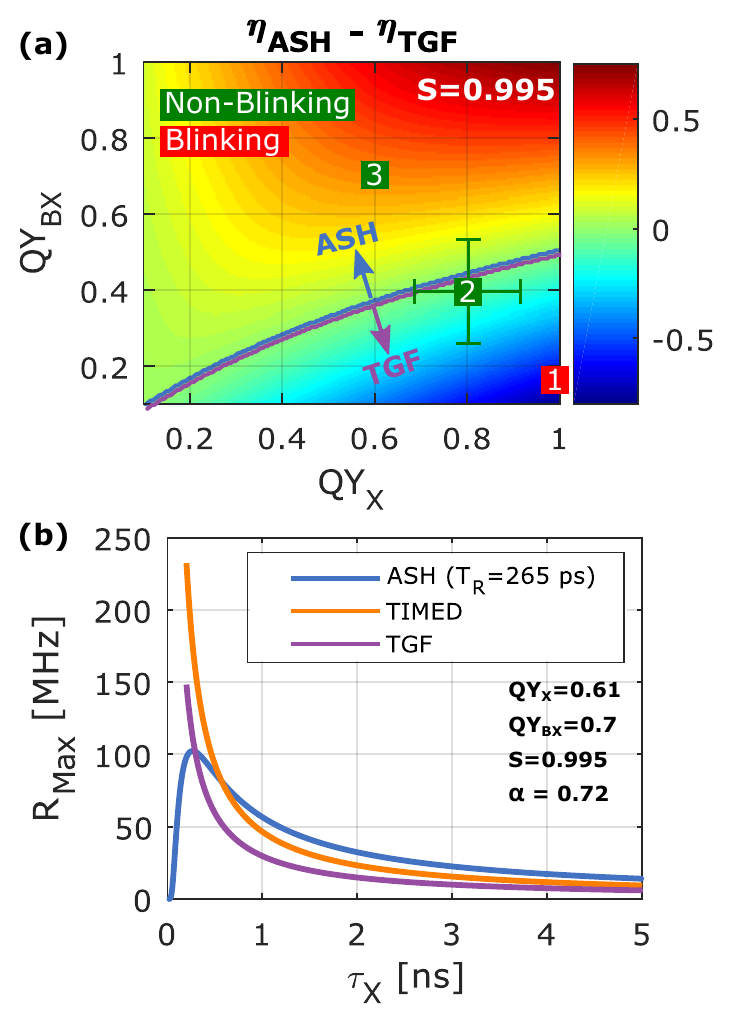}}
\caption{ (a) Difference between the efficiency of the ASH and TGF scheme displaying the regions where each technique is superior. The overlaying points are measured values of the QY$_X$ and QY$_{BX}$ for: (1) Standalone CdSe/CdZnS NQD \cite{Nair2011BiexcitonStatistics} , (2) CdSe/ 16 monolayer CdS gQD coupled to plasmonic nanocone \cite{Matsuzaki2017StrongAntenna}, and (3) CdSe/ 19 monolayer CdS standalone gQD \cite{Park2011Near-UnitySpectroscopy}. (b) The single photon rate of ASH, TIMED and TGF as a function of the exciton lifetime for $QY_X$=0.61, $QY_{BX}$=0.7, $S$=0.995, $\alpha$=0.72 and assuming an excitation rate equal to 3/$\tau_X$. }
\label{fig:comparasion}
\end{figure*}
%\begin{table}
% \begin{threeparttable}
%\caption{Comparison between a projected NQD-on-antenna system operated with TIMED and ASH photon purification protocols and other state-of-the-art SPS sources}
%\label{tab: comparasion}
%\begin{tabular}{|p{ 4.6 cm}|m{ 1.2 cm}|m{ 1.2 cm}|}
%\cline{2-3} 
%\multicolumn{1}{c|}{} & $\eta$ & S\tabularnewline
%\hline 
%NQD on Nanoantenna+TIMED, room Temperature $\left(T_{C}^{opt}\right)$&  0.15 & $>0.995$
%\tabularnewline
%\hline 
%NQD on Nanoantenna+ASH, room Temperature $\left(T_{R}=0.05 \ \tau_X \right)$ & 0.21 & $>0.995$\tabularnewline
%\hline 
%NQD on Nanoantenna+TGF, room Temperature  & 0.067 & $>0.995$\tabularnewline
%\hline 
%Parametric downconversion - nonlinear crystal\cite{Ngah2015Ultra-fastTechnology} &  $2.1\times10^{-4}$ & 0.989\tabularnewline
%\hline 
%Self-assembled QD in a microcavity, cryogenic temperature\cite{Ding2016}& 0.33 & 0.991\tabularnewline
%\hline 
%\end{tabular}
% \begin{tablenotes}
% \small
% \item $^a$ Projected values based on parameters defined in text.
% \item $^b$ Operated at 4.5-10 K.
% \item $^c$ Objective coll. eff. (NA=0.68) and polarizer (50\%) to suppress laser scattering.
% \end{tablenotes}
% \end{threeparttable}
%\end{table}
As can be seen from the model and the emulation experiments, the heralded purification of single photons from BX emission in NQDs can significantly increase their single photon purity without compromising their efficiencies, which can make such NQD sources (or other sources that have high quantum yield for two-photon emission), good candidates for high-quality room-temperature SPSs.
To demonstrate the significant superiority of the heralded techniques over TGF we plot in Fig. \ref{fig:comparasion}a the difference in the efficiency between the ASH and TGF techniques for a specific chosen purity (S=0.995). 
This displays that it is not necessary to have very high QYs for the ASH technique to outperform the TGF technique. 
In fact, currently available NQD systems already are in the regime where ASH is better. 
Importantly, as the quality of these QDs improve, and as larger and larger Purcell factors are achieved to increase their single photon brightness, it is expected that the values of the $QY$s would increase making these techniques even more advantageous. 

In order to assert the applicability of our techniques and to make a quantitative comparison with other state-of-the-art devices we calculate the expected performance of our techniques based on a model system that was previously introduced, namely an NQD coupled to the near field of a plasmonic nanocone. 
In \citet{Matsuzaki2017StrongAntenna} it has been shown experimentally that this coupling induces a very large Purcell effect, leading to a significant reduction in the radiative lifetimes of both the X and the BX, and in a very large increase of $QY_{BX}$ yielding: $QY_X=0.61$, $QY_{BX}=0.7$, $\tau_X=1.6$ ns, and $\tau_{BX}=0.5$ ns. This makes the coupled NQD a rather good bi-photon emitter.  
In addition we assume a realistic photon collection efficiency of 0.8 \cite{Abudayyeh2017QuantumSources} and a detection efficiency of 90\% \cite{Ma2015SimulationDiode,Zang2017SiliconTrapping}(i.e. $\alpha = 0.72$).

One particularly important point to consider is that for ASH the response time should be fast compared to the X lifetime. 
In the supplementary information \cite{supp} we show that an on-chip implementation of ASH could realistically have response times of about 265 ps. 
In the model system suggested in the previous paragraph this would correspond to $T_R/\tau_X \approx 0.17$ and $\eta_{ASH}=0.19$ compared to $\eta_{TIMED}=0.15$ and $\eta_{TGF}=0.089$.
Under such efficiencies the single photon rate can be estimated to be $R_{ASH}= 38$ MHz, $R_{TIMED}=30$ MHz, and $R_{TGF}=18$ MHz under pulsed excitation at a rate of 200 MHz ($1/3\tau_X$).
This is another confirmation that under realistic conditions both heralded techniques can outperform TGF.
Nonetheless it is clear that $T_R$ imposes a limitation on the maximum X decay rate that can be achieved before TIMED and TGF outperform ASH.
We analyze this situation in Fig.\ref{fig:comparasion}b where we vary $\tau_X$ while holding all other parameters constant as in the suggested model system above.
As expected at low $\tau_X$ ASH suffers however TIMED (TGF) only has better performance for $\tau_X <$0.6 (0.3) ns. 
These values of single photon purity and rate also compare well with state-of-the-art parametric heralded sources \cite{Ngah2015Ultra-fastTechnology,Wang2016ExperimentalEntanglement,Joshi2018FrequencySources}, and cryogenic-operated self-assembled QD \cite{Ding2016,Somaschi2016NearState} sources.
We note that in contrast to cryogenic QD sources, photons emitted from room-temperature sources are distinguishable in nature. However several recent reports have addressed this issue by proposing cavity enhanced techniques to increase the indistinguishablity of room-temperature emitters \cite{Grange2015,Wein2018FeasibilityCavities,Choi2018HighlyRegime}. 

One interesting application, even under limited indistinguishability, is quantum key distribution (QKD) for which true room-temperature SPSs are desired \cite{Takemoto2015QuantumDetectors}.
%We show in the supplementary information \cite{supp} that using the parameters in table \ref{tab: comparasion} the projected NQD based SPSs will be able to compete with an ideal decoy state protocol \cite{Hwang2003QuantumCommunication,Lo2005DecoyDistribution}.
In the quantum metrology realm one may devise using these highly intensity-squeezed sources to conduct weak absorption measurements on highly sensitive samples or for calibrating photodetectors with high precision well beyond the shot-noise limit \cite{Lounis2005}. 
Eventually such sources may become the ultimate standard for intensity measurements in what is known as the quantum candela \cite{Cheung2007TheRadiation}.
Furthermore the heralding nature of the schemes proposed in this work open up the possibility of multiplexing a few NQDs together to overcome any inefficiencies that may be caused due to inadequate quantum yields or collection efficiencies, thus increasing the attainable single photon rates \cite{supp}.

In summary, we show both theoretically and by experimental emulation the ability to break the trade-off between stability, efficiency and single photon purity in statistical two-photon emitters by temporal heralding schemes, allowing to achieve near-unity efficiency and single photon purity simultaneously.

\begin{acknowledgement}
This work was performed, in parts under the financial support from: The Einstein Foundation Berlin; The U.S. Department of Energy: Office of Basic Energy Sciences - Division of Materials Sciences and Engineering; User Project 2017BU0062 at the Center for Integrated Nanotechnologies, an U.S. Department of Energy (DOE) Office of Science User Facility operated for the DOE by Los Alamos National Laboratory (LANL) (Contract DE-AC52-06NA25396) and Sandia National Laboratories (Contract DE-NA-0003525). J.H. and S.M. acknowledge partial support from the LANL Laboratory Directed Research and Development funds.

\end{acknowledgement}

%%%%%%%%%%%%%%%%%%%%%%%%%%%%%%%%%%%%%%%%%%%%%%%%%%%%%%%%%%%%%%%%%%%%%
%% The same is true for Supporting Information, which should use the
%% suppinfo environment.
%%%%%%%%%%%%%%%%%%%%%%%%%%%%%%%%%%%%%%%%%%%%%%%%%%%%%%%%%%%%%%%%%%%%%
\begin{suppinfo}

% % A listing of the contents of each file supplied as Supporting Information
% % should be included. For instructions on what should be included in the
% % Supporting Information as well as how to prepare this material for
% % publications, refer to the journal's Instructions for Authors.

The supplementary information contains the theoretical derivations for the equations shown in the text in addition to experimental and analysis details.% and an assessment of the utility of the sources proposed in this work to QKD applications. 
% % \begin{itemize}
% %   \item Filename: brief description
% %   \item Filename: brief description
% % \end{itemize}

\end{suppinfo}
% \nolinenumbers
%%%%%%%%%%%%%%%%%%%%%%%%%%%%%%%%%%%%%%%%%%%%%%%%%%%%%%%%%%%%%%%%%%%%%%
%%                           REFERENCES    							%%
%%%%%%%%%%%%%%%%%%%%%%%%%%%%%%%%%%%%%%%%%%%%%%%%%%%%%%%%%%%%%%%%%%%%%%
\bibliography{Mendeley}

\begin{mcitethebibliography}{54}
\bibitem[Not()]{Note-2}
In this analysis we assume that the NQD is pumped well above saturation with a
  short pulse so as to prepare the NQD in the BX state at the beginning of each
  excitation cycle.\relax
\mciteBstWouldAddEndPunctfalse
\mciteSetBstMidEndSepPunct{\mcitedefaultmidpunct}
{}{\mcitedefaultseppunct}\relax
\EndOfBibitem
\end{mcitethebibliography}

\end{document}

\title{{\bf Supplementary Information} \protect\\ Purification of single photons by temporal heralding of quantum dot sources}

\author{Hamza Abudayyeh}%
\author{Boaz Lubotzky}
\author{Ronen Rapaport}
 \email{ronenr@phys.huji.ac.il}
\affiliation{%
Racah Institute for Physics and Center for Nanoscience and Nanotechnology, The Hebrew University of Jerusalem, Jerusalem 9190401, Israel
}
\author{Somak Majumder}
\author{Jennifer A. Hollingsworth}
\affiliation{Materials Physics \& Applications Division: Center for Integrated Nanotechnologies, Los Alamos National Laboratory, Los Alamos, New Mexico 87545, United States}

\maketitle

\section{Theoretical derivation of heralding parameters}

The quantum yield of an emission process $i$ is given by the ratio
of the radiative decay rate $\Gamma_{i}^{r}$ to total decay rate
$\Gamma_{i}$:

\[
QY_{i}=\frac{\Gamma_{i}^{r}}{\Gamma_{i}}
\]

For a biexciton-exciton emission cascade simple statistical scaling
implies that the radiative decay rate of the biexciton should be 4
times that of the exciton. In reality other factors play a role such
as the degree of confinement of the electrons and holes to the core
of a core/shell quantum dot for example. Therefore in general this
scaling factor will be called $\beta$, i.e. $\Gamma_{BX}^{r}=\beta\Gamma_{X}^{r}$
and thus:

\[
\begin{array}{ccc}
\frac{QY_{BX}}{QY_{X}} & = & \frac{\Gamma_{BX}^{r}\Gamma_{X}}{\Gamma_{BX}\Gamma_{X}^{r}}\\
 & = & \frac{\beta\Gamma_{X}^{r}\Gamma_{X}}{\Gamma_{BX}\Gamma_{X}^{r}}\\
 & = & \beta\frac{\Gamma_{X}}{\Gamma_{BX}}
\end{array}
\]

Therefore if the $\beta$-scaling factor is known a simple lifetime
measurement can be used to find the ratio of the biexciton to exciton
quantum yield. 

We will assume that our nanocrystal quantum dot can emit at most two
photons based on the biexciton-exciton cascade and that it is being
pumped well above saturation. The probability that a biexciton photon
is emitted $T_{BX}$ after the laser pulse is given by:

\begin{equation*}
p(T_{BX})=\alpha QY_{BX}\Gamma_{BX}e^{-\Gamma_{BX}T_{BX}}
\end{equation*}

whereas the exciton emission time ($T_{X}$) is conditioned on the
emission time of the biexciton as follows:

\begin{equation*}
p\left(T_{x}\right|\left.T_{BX}\right)=\alpha QY_{X}\Gamma_{X}e^{-\Gamma_{X}\left(T_{X}-T_{BX}\right)}
\end{equation*}

\begin{tabular}{llp{13 cm}l}
\multicolumn{2}{l}{where:} & \tabularnewline
 & $\alpha$: & overall detection efficiency of the system\tabularnewline
 & $QY_{i}$: & quantum yield of the $i^{\text{th}}$ emission process\tabularnewline
 & $p(T_{BX})$: & probability of biexciton event at local time $T_{BX}$\tabularnewline
 & $p\left(T_{x}\right|\left.T_{BX}\right):$ &{conditional probability of exciton event at local time $T_{X}$ given
that a biexciton event was detected at $T_{BX}$}\tabularnewline
\end{tabular}

Therefore for a standalone NQD the probability to emit one and two photons ($P_1$ and $P_2$ respectively) are:
\begin{equation*}
P_1=\alpha QY_X + \alpha QY_{BX} - 2 \alpha^2 QY_X QY_{BX} 
\end{equation*}
\begin{equation*}
P_2=\alpha^2 QY_X QY_{BX} 
\end{equation*}
\subsection{Fixed gate techniques}

If we assume that a fixed time gate is applied as in the case of TGF
and TIMED than we can ask what is the probability that a photon will
take a certain route. For example what is the probability that one
photon will arrive within $T$ from the laser pulse and the other
photon will arrive after $T$. This probability is given by the following
integral:

\begin{eqnarray*}
\mathscr{\mathcal{\mathscr{P}}}_{1}(T) & = & \alpha^{2}QY_{X}QY_{BX}\int_{0}^{T}dT_{BX}\int_{T}^{\infty}dT_{X}\,p(T_{BX})\,p\left(T_{x}\right|\left.T_{BX}\right)\nonumber \\
 & = & \alpha^{2}QY_{X}QY_{BX}\frac{\Gamma_{BX}}{\Gamma_{X}-\Gamma_{BX}}\exp\left(-\Gamma_{X}T\right)\left\{ 1-\exp\left(-\left(\Gamma_{BX}-\Gamma_{X}\right)T\right)\right\} 
\end{eqnarray*}

On the other hand the probability that both will arrive prior to $T$
$\left(\mathscr{\mathcal{\mathscr{P}}}_{2}(T)\right)$ or after $T$
$\left(\mathscr{\mathcal{\mathscr{P}}}_{3}(T)\right)$ are given by
the following integrals:

\[
\mathscr{\mathcal{\mathscr{P}}}_{2}(T)=\alpha^{2}QY_{X}QY_{BX}\int_{0}^{T}dT_{BX}\int_{T_{BX}}^{T}dT_{X}\,p(T_{BX})\,p\left(T_{x}\right|\left.T_{BX}\right)
\]

\[
\mathscr{\mathcal{\mathscr{P}}}_{3}(T)=\alpha^{2}QY_{X}QY_{BX}\int_{T}^{\infty}dT_{BX}\int_{T_{BX}}^{\infty}dT_{X}\,p(T_{BX})\,p\left(T_{x}\right|\left.T_{BX}\right)
\]

These events in addition to all other possibilities  are summarized in
table \ref{table: emission_paths}.
\begin{table*}
\caption{Possible events resulting from an emitter that has a maximum of a
two photon cascade by applying a fixed time gate\label{table: emission_paths}}
% \begin{onehalfspace}
\begin{centering}
\resizebox{\textwidth}{!}{%
\begin{tabular}{|c|ccl|}
\cline{2-4} 
\multicolumn{1}{c|}{\#} & Condition on BX & Condition on X & Prob. of Process\tabularnewline
\hline 
1. & $T_{BX}\le T$ & $T_{X}\ge T$ & $\mathscr{\mathcal{\mathscr{P}}}_{1}(T)=\alpha^{2}QY_{X}QY_{BX}\frac{\Gamma_{BX}}{\Gamma_{BX}-\Gamma_{X}}\left\{ \exp\left(-\Gamma_{X}T\right)-\exp\left(-\Gamma_{BX}T\right)\right\} $\tabularnewline
\hline 
2. & $T_{BX}\le T$ & $T_{X}\le T$ & $\mathscr{\mathcal{\mathscr{P}}}_{2}(T)=\alpha^{2}QY_{BX}QY_{x}\left\{ 1-\frac{\Gamma_{BX}}{\Gamma_{BX}-\Gamma_{X}}\exp\left(-\Gamma_{X}T\right)+\frac{\Gamma_{X}}{\Gamma_{BX}-\Gamma_{X}}\exp\left(-\Gamma_{BX}T\right)\right\} $\tabularnewline
\hline 
3. & $T_{BX}\ge T$ & $T_{X}\ge T$ & $\mathscr{\mathcal{\mathscr{P}}}_{3}(T)=\alpha^{2}QY_{BX}QY_{x}\exp\left(-\Gamma_{BX}T\right)$\tabularnewline
\hline 
4. & $T_{BX}\le T$ & \textemdash{} & $\mathscr{\mathcal{\mathscr{P}}}_{4}(T)=\alpha QY_{BX}(1-\alpha QY_{X})\left\{ 1-\exp\left(-\Gamma_{BX}T\right)\right\} $\tabularnewline
\hline 
5. & $T_{BX}\ge T$ & \textemdash{} & $\mathscr{\mathcal{\mathscr{P}}}_{5}(T)=\alpha QY_{BX}(1-\alpha QY_{X})\exp\left(-\Gamma_{BX}T\right)$\tabularnewline
\hline 
6. & \textemdash{} & $T_{X}\le T$ & $\mathscr{\mathcal{\mathscr{P}}}_{6}(T)=\alpha QY_{X}(1-\alpha QY_{BX})\left\{ 1-\exp\left(-\Gamma_{X}T\right)\right\} $\tabularnewline
\hline 
7. & \textemdash{} & $T_{X}\ge T$ & $\mathscr{\mathcal{\mathscr{P}}}_{7}(T)=\alpha QY_{X}(1-\alpha QY_{BX})\exp\left(-\Gamma_{X}T\right)$\tabularnewline
\hline 
8. & \textemdash{} & \textemdash{} & $\mathscr{\mathcal{\mathscr{P}}}_{8}=\left(1-\alpha QY_{BX}\right)\left(1-\alpha QY_{X}\right)$\tabularnewline
\hline 
\end{tabular}} 
\par\end{centering}
% \end{onehalfspace}

\end{table*}

In Time Gated Filtering (TGF) shown in figure \ref{fig: TGF}a the switch is open to an optical dump from the beginning of each pulse up to a filtering time $T_F$ after which the photons are routed to the signal port.
In such a manner the short lifetime components of an optical signal can be filtered out.
This can be used to filter out the biexciton emission where the  the efficiency of the source
would be the probability of only one photon events after a time gate
of $T_{F}$ namely:

\begin{eqnarray*}
\eta_{TGF} & = & \mathscr{\mathcal{\mathscr{P}}}_{1}(T_{F})+\mathscr{\mathcal{\mathscr{P}}}_{5}(T_{F})+\mathscr{\mathcal{\mathscr{P}}}_{7}(T_{F})
\end{eqnarray*}

where as the purity is the same quantity normalized by the probability
of obtaining one or two photons after $T_{F}$ which is given by:

\begin{eqnarray*}
S_{TGF} & = & \frac{\mathscr{\mathcal{\mathscr{P}}}_{1}(T_{F})+\mathscr{\mathcal{\mathscr{P}}}_{5}(T_{F})+\mathscr{\mathcal{\mathscr{P}}}_{7}(T_{F})}{\mathscr{\mathcal{\mathscr{P}}}_{1}(T_{F})+\mathscr{\mathcal{\mathscr{P}}}_{3}(T_{F})+\mathscr{\mathcal{\mathscr{P}}}_{5}(T_{F})+\mathscr{\mathcal{\mathscr{P}}}_{7}(T_{F})}
\label{eq: S_TGF}
\end{eqnarray*}
This case of TGF with QY$_X$=0.6 and QY$_{BX}$=0.71 \cite{Matsuzaki2017StrongAntenna} is plotted in figure \ref{fig: TGF}b where the trade-off between purity and efficiency in terms of $T_F$ is clear. 

For the special case of unity collection efficiency and unity quantum
yields the expressions reduce to:

\begin{eqnarray*}
\eta_{TGF}^{ideal} & = & \frac{\Gamma_{BX}}{\Gamma_{BX}-\Gamma_{X}}\left\{ \exp\left(-\Gamma_{X}T_{F}\right)-\exp\left(-\Gamma_{BX}T_{F}\right)\right\} 
\end{eqnarray*}

\begin{eqnarray*}
S_{TGF}^{ideal} & = & \frac{\Gamma_{BX}\left\{ \exp\left(-\Gamma_{X}T_{F}\right)-\exp\left(-\Gamma_{BX}T_{F}\right)\right\} }{\Gamma_{BX}\exp\left(-\Gamma_{X}T_{F}\right)-\Gamma_{X}\exp\left(-\Gamma_{BX}T_{F}\right)}
\end{eqnarray*}

% Note that as in standalone NQDs the efficiency of the source also
% determines its determinicity (how well one can predict the desired
% event).

On the other hand for the TIMe resolved heraldED (TIMED) scheme only
the first process constitutes a successful heralding event namely:

\begin{eqnarray*}
\eta_{TIMED} & = & \mathscr{\mathcal{\mathscr{P}}}_{1}(T_{C})=\alpha^{2}QY_{X}QY_{BX}\frac{\Gamma_{BX}}{\Gamma_{BX}-\Gamma_{X}}\left\{ \exp\left(-\Gamma_{X}T_{C}\right)-\exp\left(-\Gamma_{BX}T_{C}\right)\right\} 
\end{eqnarray*}

By using $\tau_{i}=\Gamma_{i}^{-1}$ for $i=X,BX$ we arrive at the
equation stated in the main text.

We can choose the optimum cutoff time $T_{c}^{opt}$ that will maximize
$\eta$. This is given in units of the exciton lifetime as:

\begin{equation*}
\Gamma_{x}T_{c}^{opt}=\left(\frac{\Gamma_{BX}}{\Gamma_{X}}-1\right)^{-1}\ln\left(\frac{\Gamma_{BX}}{\Gamma_{X}}\right)\label{eq:tcopt}
\end{equation*}

Using this the optimum efficiency $\eta^{opt}$ is :

\begin{equation*}
\eta_{TIMED}^{opt}=\alpha^{2}QY_{X}QY_{BX}\frac{\nicefrac{\Gamma_{BX}}{\Gamma_{X}}}{\nicefrac{\Gamma_{BX}}{\Gamma_{X}}-1}\left\{ \left(\frac{\Gamma_{BX}}{\Gamma_{X}}\right)^{\frac{-1}{\nicefrac{\Gamma_{BX}}{\Gamma_{X}}-1}}-\left(\frac{\Gamma_{BX}}{\Gamma_{X}}\right)^{\frac{-\nicefrac{\Gamma_{BX}}{\Gamma_{X}}}{\nicefrac{\Gamma_{BX}}{\Gamma_{X}}-1}}\right\} \label{eq:etaopt}
\end{equation*}

An important parameter for the case
of multiplexed sources is the determinicity defined as the ratio of
true heralded events to overall trigger events. This parameter defines
the reliability of the trigger signal in heralding a signal photon.
In statistics this often called the Positive Predictive Value (PPV)
defined as the ratio of true postives to overall postives, but we
will stick to the term determinicity for comparasion with non-heralded
sources. For the TIMED scheme the determinicity $D$ is given by:

\begin{equation*}
D_{TIMED}=\frac{\mathscr{\mathcal{\mathscr{P}}}_{1}(T_{C})}{\mathscr{\mathcal{\mathscr{P}}}_{1}(T_{C})+\mathscr{\mathcal{\mathscr{P}}}_{2}(T_{C})+\mathscr{\mathcal{\mathscr{P}}}_{4}(T_{C})+\mathscr{\mathcal{\mathscr{P}}}_{6}(T_{C})}
\end{equation*}

Again for the idealistic case of unity collection efficiency and quantum
yields this reduces to:

\begin{equation*}
D_{TIMED}^{ideal}=\frac{\Gamma_{BX}}{\Gamma_{BX}-\Gamma_{X}}\frac{\exp\left(-\Gamma_{X}T_{C}\right)-\exp\left(-\Gamma_{BX}T_{C}\right)}{1-\exp\left(-\Gamma_{BX}T\right)}
\end{equation*}

\subsection{Active Switching Heralded (ASH) scheme}

Due to the active switching the only parameter of importance for this
scheme is the resolution time $T_{R}$ of the system. It is the difference
in the arrival time between the two photons that needs to be taken
into consideration. There are five possible outcomes:
\begin{itemize}
\item Two photons with temporal seperation more than $T_{R}$ (successful
heralding event) : 
\begin{eqnarray*}
\mathscr{\mathcal{\mathscr{P}}}_{2S} & = & \alpha^{2}QY_{X}QY_{BX}\int_{0}^{\infty}dT_{BX}\int_{T_{BX}+T_{R}}^{\infty}dT_{X}\,p(T_{BX})\,p\left(T_{x}\right|\left.T_{BX}\right)\\
 & = & \alpha^{2}QY_{X}QY_{BX}\exp\left(-\Gamma_{X}T_{R}\right)
\end{eqnarray*}
\item Two photons with temporal seperation arriving within $T_{R}$ from
each other:
\begin{eqnarray*}
\mathscr{\mathcal{\mathscr{P}}}_{2F} & = & \alpha^{2}QY_{X}QY_{BX}\int_{0}^{\infty}dT_{BX}\int_{T_{BX}}^{T_{BX}+T_{R}}dT_{X}\,p(T_{BX})\,p\left(T_{x}\right|\left.T_{BX}\right)\\
 & = & \alpha^{2}QY_{X}QY_{BX}\left\{ 1-\exp\left(-\Gamma_{X}T_{R}\right)\right\} 
\end{eqnarray*}
\item Exciton photon emission only:
\begin{eqnarray*}
\mathscr{\mathcal{\mathscr{P}}}_{1X} & = & \alpha QY_{X}\left(1-\alpha QY_{BX}\right)
\end{eqnarray*}
\item Biexciton photon emission only: 
\begin{eqnarray*}
\mathscr{\mathcal{\mathscr{P}}}_{1BX} & = & \alpha QY_{BX}\left(1-\alpha QY_{X}\right)
\end{eqnarray*}
\item No photon emission: 
\begin{eqnarray*}
\mathscr{\mathcal{\mathscr{P}}}_{0} & = & \left(1-\alpha QY_{X}\right)\left(1-\alpha QY_{BX}\right)
\end{eqnarray*}
\end{itemize}
Only the first of these events is a successful heralding event therefore
the efficiency is given by:

\begin{equation*}
\eta_{ASH}=\alpha^{2}QY_{X}QY_{BX}\exp\left(-\Gamma_{X}T_{R}\right)
\end{equation*}

On the other hand all but the last of these outcomes still will consitute
a trigger event therefore the determinicity can be written as: 

\begin{eqnarray*}
D_{ASH} & = & \frac{\mathscr{\mathcal{\mathscr{P}}}_{2S}}{\mathscr{\mathcal{\mathscr{P}}}_{2S}+\mathscr{\mathcal{\mathscr{P}}}_{2F}+\mathscr{\mathcal{\mathscr{P}}}_{1X}+\mathscr{\mathcal{\mathscr{P}}}_{1BX}}\nonumber \\
 & = & \frac{\alpha^{2}QY_{X}QY_{BX}\exp\left(-\Gamma_{X}T_{R}\right)}{\alpha QY_{X}+\alpha QY_{BX}+\alpha^{2}QY_{X}QY_{BX}}
\end{eqnarray*}

\begin{figure}
\begin{centering}
\includegraphics{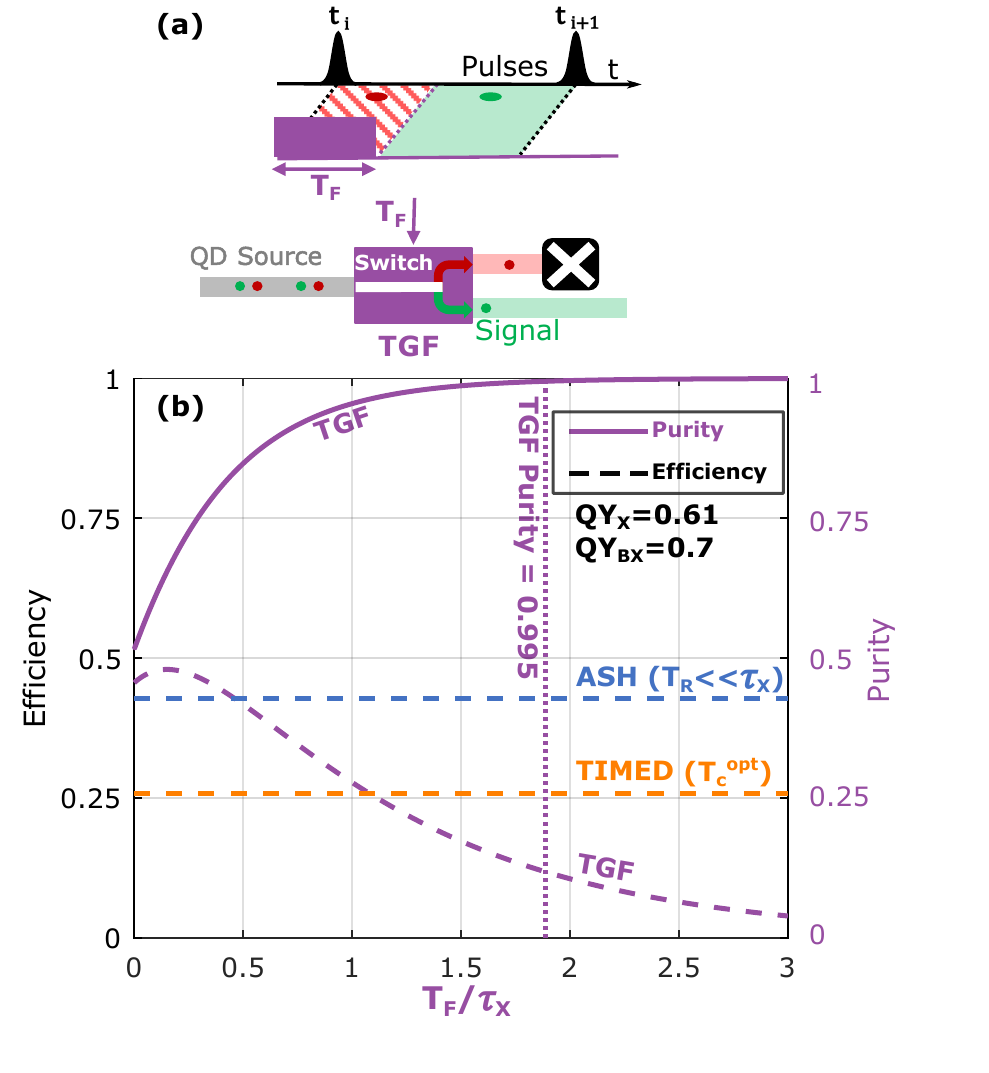}
\par\end{centering}
\caption{(a) Schematic of Time-Gated Filtering technique (TGF). (b) The purity (solid line) and efficiency (dashed lines) of the TGF technique for the quantum yields corresponding to a gQD coupled to a nanocone \cite{Matsuzaki2017StrongAntenna}. This is compared to the efficiency of the ASH and TIMED technique \label{fig: TGF}}
\end{figure}

\begin{figure}
\begin{centering}
\includegraphics{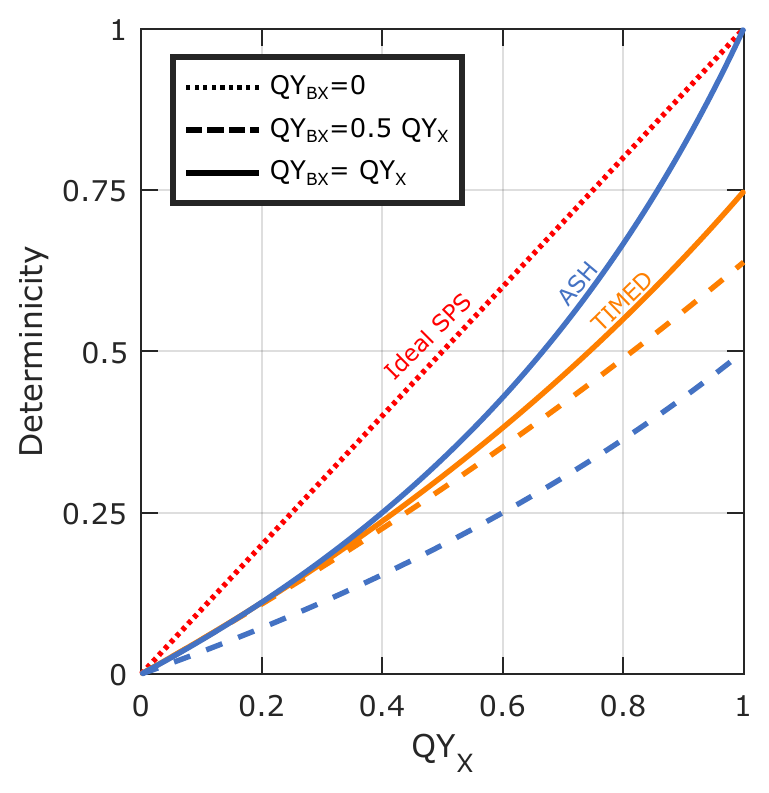}
\par\end{centering}
\caption{Determinicity as a function of the quantum yield for ASH ($T_{R}<<\tau_X)$
and TIMED ($T_{C}^{opt})$ as compared to an ideal SPS \label{fig:Determinicity}}
\end{figure}

To establish the relevance of these techniques we compare the efficiency of ASH and TIMED to the simpler TGF technique for the case of $QY_X=0.61$ and $QY_{BX}=0.7$ (see main text) \cite{Matsuzaki2017StrongAntenna} in figure \ref{fig: TGF}b. 
For TGF $T_F>1.8$ (indicated by the dotted line) is needed to produce the same purity (0.995)  measured in our heralded experiments.
For this value of $T_F$ it is clear that the efficiency of both ASH and TIMED are superior to TGF.

Figure \ref{fig:Determinicity} displays the determinicity as a function
of the quantum yield for equal quantum yields (solid line) and for
the case where $QY_{BX}=0.5QY_{X}$ (dashed line). As can be clearly
seen the ASH scheme has a determinicity of unity for unity quantum
yields but TIMED reaches to a maximum of around 75\% when the optimum
cutoff time is used. It should be noted however that this cutoff time
was optimized for efficiency not determinicity and in principle one
can operate at higher determinicities at the cost of lower efficiency.
In the non-ideal case $\left(QY_{BX}=0.5QY_{X}\right)$ the TIMED
scheme overcomes the ASH scheme since in this case false positives
start to appear. ASH is more susceptible to false triggers since it
is always open to the idler port until a trigger photon has arrived
, whereas the TIMED scheme switches after $T_{C}$ regardless of the
presence of a trigger photon or not.

Determinicity is an important parameter for a multiplexed heralded source.
This parameter is important because it defines the upper limit for the overall efficiency of the multiplexed source since the optical router will rely on the trigger signal to switch between different components.
The schemes proposed in this paper have maximum theoretical determinicities that approach unity for optimized parameters.

\section{Experimental Details}

\subsection{Setup and data format }

The measurement is conducted by spin coating a sample of core/thick
shell CdSe/CdS nanocrystal quantum dots dispersed in PMMA onto a silicon
wafer. The input laser light is a femtosecond 2 MHz 405 nm pulsed
laser generated by second harmonic generation of a femtosecond Ti:Sapphire
laser operating at 810 nm. The sample is scanned using a periscopic
system of scanning stages and the laser light is focused onto the
sample using a 0.9 NA objective (Olympus MPLFLN100xBD). The emission from a single NQD
is collected using the same objective and directed to the collection
arm using a 600nm shortpass dichroic mirror. The sample is then imaged
using a CMOS camera to verify the excitation of a single NQD. After that the emission is redirected to a set of beam-splitters and single
photon avalanche photodiodes (Excelitas SPCM-AQRH-14-FC) as shown in figure 3a in the
main text. 

The signal from each detector is routed to a different channel on
the timetagger device (Swabian TimeTagger 2.0). The output from the timetagger to the computer
is a vector containing the arrival times of all the photons within
the set exposure time with respect to the beginning of the measurement
in addition to a tag labeling which channel this count came from.
These arrival times are commonly referred to as global times. Another
channel on the time-tagger records the excitation pulse times. By
comparing the global time of each count with the nearest preceding
laser pulse one can find the local time for the count. A histogram
of these local times is what constitutes a lifetime measurement. Therefore
at the end of this step we have information about which pulse and
what channel the count came from, and its global and local times.
This constitutes all the information needed for subsequent analysis
steps.

\subsection{Photon number correction}

Due to the deadtime of each detector, counts from the same detector
within the same pulse must be neglected. If this was not conducted
then a bias would be introduced since two photons arriving with a
time separation less than the dead time would not be detected whereas
photons separated by more than the deadtime would be. However this
does introduce an error in estimating the two and three photon counts
due to the finite probability that two or more photons would
arrive at the same detector. To correct for this error we must the
calculate the probability for the photons to arrive at different detectors
and correct the photon counts by dividing by this probability. We
have a two beamsplitter scheme the first with a reflectivity $R_{1}$ and
the second with a reflectivity $R_{2}$. Assuming two photons were
emitted the probability that they would arrive at different detectors
is a simple exercise in probabilities and is given by:

\[
P_{diff}^{(2)}=2R_{1}\left(1-R_{1}\right)+\left(1-R_{1}\right)^{2}\left[2R_{2}\left(1-R_{2}\right)\right]
\]
If the pulse contains three photons the probability of them each arriving
at a different detector is given by:

\[
P_{diff}^{(3)}=3R_{1}\left(1-R_{1}\right)^{2}\left[2R_{2}\left(1-R_{2}\right)\right]=6R_{1}R_{2}\left(1-R_{1}\right)^{2}\left(1-R_{2}\right)
\]
Therefore the ``true'' photon numbers ($N_{2}$ and $N_{3}$) are
related to the measured ones ($N_{2m}$ and $N_{3m}$) by:

\begin{eqnarray*}
N_{2} & = & \frac{N_{2m}}{P_{diff}^{(2)}}\\
N_{3} & = & \frac{N_{3m}}{P_{diff}^{(3)}}
\end{eqnarray*}
In our case $R_{1}=0.4$ , $R_{2}=0.5$ and therefore:

\begin{eqnarray*}
N_{2} & = & 1.52\,N_{2m}\\
N_{3} & = & 4.63\,N_{3m}
\end{eqnarray*}
From now on it will be assumed that this correction is conducted whenever
there is a two photon or three photon event. 

\subsection{Analysis Method}

% \subsubsection*{TGF}

% Using the data as set before we throw out all counts with local times
% less than $T_{F}$ and with the remaining data we do the following:
% \begin{itemize}
% \item A correlation measurement on the remaining global data to obtain the
% second order correlation function.
% \item Calculate the number of two photon $\left(N_{2}\right)$ and one photon
% $\left(N_{1}\right)$ events and calculate the purity using $S=1-\nicefrac{N_{2}}{\alpha N_{1}}$.
% The factor of $\alpha$ is present because the probability of collecting
% two photons emitted by the NQD is $\alpha^{2}$ where the probability
% to collect a single photon is $\alpha$. Since we are interested in
% the purity of the photons emitted from the source rather than the
% photons seen in the detectors we must normalize by this factor. Furthermore
% we make two more justifiable assumptions in this formula. First we
% assume that the 3 photon probability is negligible and indeed the three photon probability
% is three orders of magnitude lower than the two photon probability.
% Also we assume that $N_{2}\ll N_{1}$ which is also the case. 
% \end{itemize}

% \subsubsection*{Heralded Schemes}
Prior to purification the purity of the QD was calculated by comparing the number of multiphoton ($N_{\ge2}$) events to single photon events ($N_{1}$) using this formula:  
$$S= \frac{N_1}{N_1+\nicefrac{N_{2}}{\alpha}+\nicefrac{N_{3}}{\alpha^2}} $$
Higher order terms were neglected due to the much lower probability of four or more emission events. This also is consistent with the ability to resolve up to three photons in our setup.   
The factor of $\alpha$ in the denominator is to account for the extra optical loss that a two photon and three photon states encounter. 

In the heralded schemes we first filter out all counts having local
times less than $T_{F}$=300 ps to reduce the effect of correlated noise.
Then all pulses containing two or more photons were chosen for further
analysis. 

For the TIMED scheme the condition implemented afterwards is that
one (or more) photons had a local time less than $T_{C}$ and one
(or more) photons had a local time more than $T_{C}$. The number
of these events constituted the number of successful heralding events
and divided by the overall number of excitation pulses yields the
heralding efficiency. This is what is
plotted in figure 3d in the main text. From these figures we chose
the $T_{C}$ that yields the maximum efficiency and for the successful
heralding events we count the number of photons in each case that
arrive after $T_{C}$ which will be called the signal photons. The
number of cases in which we have two signal photons per heralding
event $N_{2}$ compared to the number of cases where there is only
one $N_{1}$ gives the purity: $S=1-\nicefrac{N_{2}}{\alpha N_{1}}$
which can be plotted as a function of the original filtering time
$T_{F}$ as in figure 3e in the main text. The factor of $\alpha$ in the denominator is to account for the extra optical loss that a two photon state encounters. 

In the ASH technique the time differences between the two (or more)
photons is calculated. If this time difference is more than $T_{R}$
then this is counted as a successful heralding event and the efficiency
is calculated as in the previous
case. Again in these heralding events the number of photons following
the trigger photon is counted ($N_{2}$ if there is 2 and $N_{1}$
if there is 1) and the purity can be calculated using the same formula
as above as a function of the filtering time. 

\subsection{Collection efficiency estimation}

To estimate the collection efficiency in our setup we take into account the efficiencies of the various optical components using three techniques:
\begin{itemize}
\item Simulation (sim) of collection efficiency into objective by using
a commercial FDTD software (Lumerical)
\item Transmission/Reflection measurement using a 655 nm diode laser (meas).
\item Factory data (fact)
\end{itemize}
The efficiency of our system is estimated in table \ref{tab: Coll_eff}. 

\begin{table}
\caption{System collection efficiency estimation}\label{tab: Coll_eff}
\begin{centering}
\begin{tabular}{|c|c|c|}
\hline 
Component & Method & Efficiency\tabularnewline
\hline 
\hline 
Collection into objective & sim & 39.0 \%\tabularnewline
\hline 
Objective transmission & meas & 90.0 \%\tabularnewline
\hline 
600 nm SP dichroic & meas & 96.6 \%\tabularnewline
\hline 
700 nm SP filter & meas & 88.5 \%\tabularnewline
\hline 
550 nm LP filter & meas & 95.2 \%\tabularnewline
\hline 
600 nm LP filter & meas & 84.3 \%\tabularnewline
\hline 
Beam splitters (2) & meas & 86.3 \%\tabularnewline
\hline 
Mirrors (6) & meas & 75.9 \%\tabularnewline
\hline 
Fiber coupling & meas & 80.0 \%\tabularnewline
\hline 
Detector efficiency & fact & 70.0 \%\tabularnewline
\hline 
\multicolumn{2}{|r|}{\textbf{Total}} & \textbf{8.83 \%}\tabularnewline
\hline 
\end{tabular}
\par\end{centering}

\end{table}

\subsection{Effect of noise on single photon purity }
Experimentally we expect that the sole source reducing $S$ in the signal port is from noise. 
To check this point we show the purity, $S$, of both schemes as a function of the filtering time $T_F$ after the excitation pulse. 
We can see from figure 3e in the main text that the ASH purity improves dramatically with even 1 ns of filtering which fits well with the extracted correlated noise lifetime. 
The TIMED scheme, on the other hand, is hardly affected by the filtering. 
This difference in sensitivity can be understood since in the TIMED scheme we apply a passive gate, so all short-lifetime counts will by default be directed to the idler port without affecting the fraction of photons arriving to the signal port. 
On the other hand, in the ASH scheme a short lifetime count will lead to premature "switching" to the signal port causing both bi-exciton and exciton photons to be directed to the signal port. 
To confirm that that the short-lifetime counts are indeed from correlated noise counts and not from higher order multiexcitons we estimate their rate based on the measured ASH purity at $T_F=0$.
This is equivalent to finding the ratio of three photon to two photon events in the raw data. 
If indeed correlated noise is the source for the reduced $T_F=0$ purity, this ratio  should give the probability per pulse to get a correlated noise count which turns out to be $1.4 \times 10^{-3}$ or correspondingly the rate is $\sim 2800$ cps. 
This agrees with the correlated noise level measured at the same excitation power on the same substrate in a region with no quantum dots ($\sim 2500-3500$ cps). 
For $T_F>1$ ns, $S$ reaches a maximum of around 0.995, after filtering out nearly all correlated noise.
Uncorrelated noise is what limits this value from reaching unity. 
Using the procedure as above we find that this noise rate is around 880 cps which is in agreement with our uncorrelated noise rate $\sim$ 600-800 cps.

To confirm these statements we will attempt to reconstruct the measured purities based on the error rates stated above.
In general the single photon purity of a signal can be defined as:
\[
S=\frac{P_{1}}{P_{\ge1}}=1-\frac{P_{\ge2}}{P_{\ge1}}
\]
where $P_{1}$, $P_{\ge1}$ and $P_{\ge2}$ are the probabilities
per relevant event of obtaining a single photon count, at least one
photon , and at least two photons respectively.

Due to the low quantum yields in our case we can assume that $P_{\ge2}\approx P_{2}$ and
therefore:

\begin{equation*}
S=1-\frac{P_{2}}{P_{\ge1}}
\end{equation*}
Note the absence of the $\alpha$ factor here since we are theoretically
considering the events obtained directly from the emitter. Now assume
we have a certain probability per pulse $\eta_{cn}=1.4\times10^{-3}/\alpha$ and $\eta_{un}=4.4\times10^{-4}/\alpha$
of obtaining correlated and uncorrelated noise event respectively. 
We will now study the effect of these noise terms on our measured purity:

\subsubsection*{Heralded schemes}

For the heralded schemes in principle the purity should be unity unless
there is noise. $P_{2}$ and $P_{\ge1}$
are the probabilities of getting two photons and at least one photon
\textbf{conditioned on the presence of another trigger photon.} For
simplicity lets consider the ASH technique with $T_{R}=0$ in two
regimes:
\begin{itemize}
\item In the presence of correlated noise ($T_{F}=0$) in this case $P_{\ge1}\approx QY_{X}QY_{BX}+\left(\eta_{cn}+\eta_{un}\right)\left(QY_{X}+QY_{BX}\right)$
notice that due to the heralding requirement this is effectively the
probability of two photons in a non heralding scheme. On the other
hand the probability of obtaining two photons in the signal port of
a heralding technique is just the correlation between the noise counts
and the heralded counts given by $P_{2}=QY_{X}QY_{BX}\left(\eta_{cn}+\eta_{un}\right)$
. Therefore by using the values obtained from our experiment we find
that the purity is given by:
\[
S_{with\,noise}=1-\frac{QY_{X}QY_{BX}\left(\eta_{cn}+\eta_{un}\right)}{QY_{X}QY_{BX}+\left(\eta_{cn}+\eta_{un}\right)\left(QY_{X}+QY_{BX}\right)}=0.9867
\]
\item In the absence of correlated noise ($T_{F}\gg\tau_{cn}$) and assuming
that $T_{F}$ is still much shorter than the biexciton lifetime then
we can effectively set $\eta_{cn}=0$ in the previous equation to
obtain:
\[
S_{with\,noise}=1-\frac{QY_{X}QY_{BX}\eta_{un}}{QY_{X}QY_{BX}+\eta_{un}\left(QY_{X}+QY_{BX}\right)}=0.996
\]
This is in good agreement with the results shown in figure 3e of the
main text. 
\end{itemize}

\begin{figure}[ht!]
\begin{centering}
\includegraphics{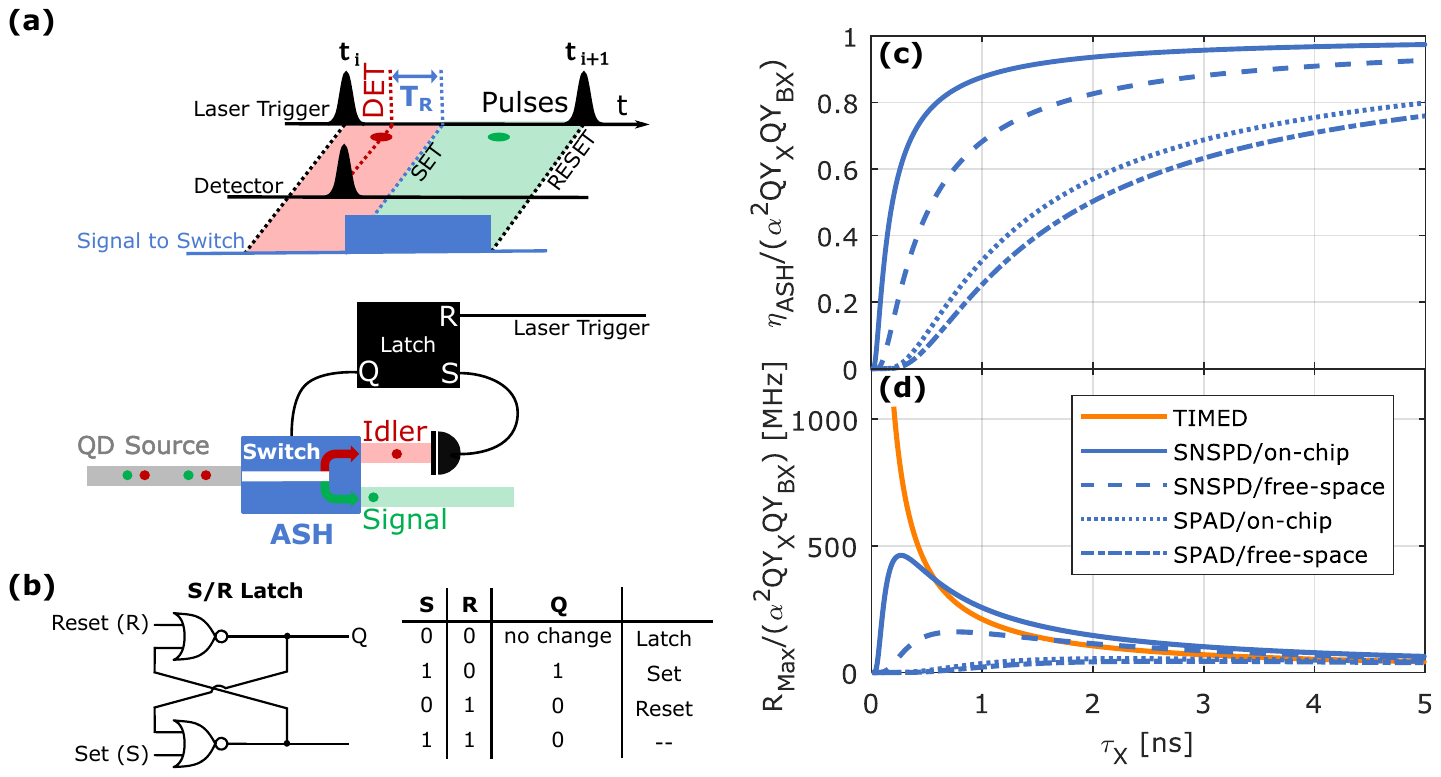}
\par\end{centering}
\caption{Schemes for the (a) ASH scheme and (b) S/R latch. (c) The heralding efficiency and (d) single photon rate of ASH under different response times as compared to the TIMED technique  \label{fig:Implementation}}
\end{figure}

\section{Practical Details}

\begin{table}[h!]
(a)

\begin{tabular}{|c|c|c|c|}
\hline 
Circuit Component & Type & Description & Value\tabularnewline
\hline 
\hline 
\multirow{4}{*}{Detector} & \multirow{2}{*}{SNSPD} & Latency & $\sim$ 50 ps \cite{Allmaras2018IntrinsicDetectors}\tabularnewline
\cline{3-4} 
 &  & Jitter & 15 ps \cite{SingleQuantumSingleEos}\tabularnewline
\cline{2-4} 
 & \multirow{2}{*}{SPAD} & Latency & $\sim$ 2 ns \cite{Prochazka2011MeasurementTransfer}\tabularnewline
\cline{3-4} 
 &  & Jitter & 50 ps \cite{PicoQuantGmbHPDMSeries}\tabularnewline
\hline 
Logic Circuit & S/R Latch & Propagation delay & 185 ps \cite{ONSemiconductor2.5V/3.3VSmartGate} \tabularnewline
\hline 
Optical Switch & 25 GHz Modulator & Rise time & 15 ps \cite{iXblueNIR-MX800-LN-10}\tabularnewline
\hline 
\multirow{2}{*}{Optical/Electronic Propagation} & On Chip &  & Negligible\tabularnewline
\cline{2-4} 
 & Free Space &  & 500 ps\tabularnewline
\hline 
\end{tabular}

(b)

\begin{tabular}{c|c|c|}
 & Free Space & On-Chip\tabularnewline
\hline 
\hline 
SNSPD & 765 ps & 265 ps\tabularnewline
\hline 
SPAD & 2.75 ns & 2.25 ns\tabularnewline
\hline 
\end{tabular}

\caption{(a) Contributions of various components to $T_R$. (b) The total response time under different configurations. \label{Tab:TR}.}
\end{table}

In this section we will discuss in detail the effect of response time on the performance of ASH.
The diagram of the proposed optical and electronic circuit needed to implement this technique is displayed \ref{fig:Implementation}a and b. We  conduct a detailed analysis of the different components that will contribute to $T_R$ in table \ref{Tab:TR} where the propagation delay of the S/R latch is the composite delay of two NOR gates \cite{ONSemiconductor2.5V/3.3VSmartGate}. 
We consider different situations with either single photon avalanche detectors (SPAD) or semiconductor nanowire single photon detectors (SNSPD) combined with the other components either on-chip or free space. We compare the efficiency and single photon rate of the resulting 4 scenarios in Fig. \ref{fig:Implementation}c and d with a comparison with TIMED.
Depending on the specific implementation the ASH technique will be better than TIMED for $\tau_X$ more than a certain value. 

\bibliography{Mendeley}
\bibliographystyle{ieeetr}